# Exploring crossmodal perceptual enhancement and integration in a sequence-reproducing task with cognitive priming

*Feng Feng, Puhong Li, Tony Stockman*

**Abstract** Leveraging the perceptual phenomenon of crossmodal correspondence has been shown to facilitate peoples information processing and improves sensorymotor performance. However, for goal-oriented interactive tasks, the question of how to enhance the perception of specific crossmodal information, and how crossmodal information integration takes place during interaction, is still unclear. The present paper reports two experiments investigating these questions. In Experiment 1, a cognitive priming technique was introduced as a way to enhance the perception of two crossmodal stimuli, the pitch-brightness and the pitch-elevation stimuli, in two conditions respectively, and their effect on sensory-motor performance was observed. Based on the results, experiment 2 combined the two crossmodal stimuli in the same interfaces in a way that their correspondence congruency was mutually exclusive. The same priming technique was applied as a manipulating factor to observe the crossmodal integration process. Results showed that first, people's crossmodal perception during interaction can be enhanced by the priming technique, but the effect varies according to the combination of crossmodal stimuli and the types of priming material. Moreover, people's subjective evaluations to-wards priming types were in contradiction with their objective behavioural data. Second, when two crossmodal sequences can be perceived simultaneously, results suggested different perceptual weights are possessed by different participants, and the perceptual enhancement effect was observed only on the dominant one (Pitch-elevation). Furthermore, the crossmodal information tended to be integrated in a selective manner with the priming technique, and in an additive manner without priming. These results contribute design implications for multisensory feedback and mindless computing.

## 1 Introduction

Understanding how people integrate multimodal information to optimise sensory-motor activity is important for multisensory system design. The multimodal perceptual phenomenon of crossmodal correspondence (CC) has long been investigated in both the cognitive behaviour and human-computer interaction (HCI) re-search fields. CC refers to the associated perceptual relationship between two or more sensory modalities, from which information perceived through visual, audio and other sensory channels are combined to form a stable output [45]. For example, an upward visual stimulus combined with an increased auditory pitch usually results in improved perceptual judgement accuracy compared

E-mail:

with one combined with decreasing pitch value. CCs have been shown to modulate people's sensory-motor reactions, and the phenomenon is shared by a large number of people [45,6] across age groups [51,55]. There is mounting empirical evidence showing that by providing external crossmodal stimuli that are varied congruently across sensory modalities, peoples sensory-motor performance can be improved [32, 13,7]. This observation has mostly resulted from studies taking a bottom-up approach, in which they have provided external stimuli, for example, a high-pitched tone paired with bright visual stimuli, enabling quicker and more accurate sensory-motor responses than that with a low-pitched tone paired with a bright visual stimuli [27,57,15]. Alongside studies that have taken the bottom-up approach, the same effect has been observed in studies taking a top-down approach, in which the perceptual association between sensory modalities depends on similar meaning or identity, such as the association of an image of a dog with a barking sound [27, 57,15].

In the HCI domain, crossmodal information has been applied in various multisensory design situations, using both bottom-up and top-down approaches, with the research emphasis on robust information perception [35,21,17], sensory substitution [29], sports skill acquisition [42,43,41] ,embodied interaction experience[4], and data representation through multisensory modalities [53,30,19]. Interestingly, some applications have been designed where congruent crossmodal associations were not always in line with intended activity. For example, the crossmodal association between red/blue colour and near/far distance were in conflict with a different type of association, i.e. red/blue map to wrong/correct activities [17]. To our knowledge, there has been relatively little research into the situation in which several crossmodal associations have co-existed in one interactive system. Thus we lack understanding of whether the perception of specific crossmodal information can be enhanced by contextual information in an interactive system, as well as how people integrate multiple crossmodal information when multiple CCs are combined either congruently or in a mutually exclusive manner. Consequently, in this paper we aim to investigate two research questions:

RQ1: Whether people's crossmodal perception can be enhanced by external contextual information, and if it can be, to what extent it modulates small and fast sensory-motor responses in a goal-oriented interactive task?

RQ2: How do people integrate crossmodal information in which two CCs are combined where the perception of the congruency of one CC excludes the perception of the congruency of the other? Furthermore, how, if at all, does the integration of such information change in the presence or absence of cognitive priming?

Cognitive priming in HCI studies is a technique to enhance cognitive performance by communicating facilitating information without overloading the cognitive channel [26,12,3]. It has been shown to have a facilitatory effect on cognitive function such as perception [3, 38], judgement [20], and affectiveness [26] during interaction. Thus we addressed the first question by adopting an existing cognitive priming technique from the HCI research field as a manipulation factor [20,24], with the purpose of engaging people with two CC pairs respectively in a top-down manner, and investigating the influence of this upon subsequent interaction behaviour. To address the second question, we conducted a second experiment by implementing the same crossmodal stimuli in a single task, but arranged in a mutually exclusive manner, and then observing how people integrate contradictory crossmodal information with and without the cognitive priming technique.

The novelty of the present study lies in the investigation of the effect of a top-down invocation of grounded crossmodal experience, seeking to expand our understanding of enhanced perception and correlated sensorymotor performance in situations with either isolated or combined crossmodal information. The first contribution is that we introduced two types of crossmodal priming material and systematically evaluate the priming effect based on two isolated CCs. The second contribution, to the best of our knowledge, is the first investigation of the crossmodal information integration process under a goal-oriented task, when the congruency of the crossmodal stimuli are mutually exclusive to perceive.

This paper starts by reviewing related work on crossmodal interaction and priming techniques used in HCI studies (section two), during which working definitions and terminology are also introduced. We then go on to explain our research motivation and questions in section three. Detailed information about the experimental procedure, task descriptions and participants is presented in section four. Sections five and six present the results and discussions of experiment 1 and 2 respectively. Finally, we discuss the overall results and draw conclusions in section seven.

## 2 Related work

### 2.1 A view from cognition research and behavioural studies

Our perception of certain physical features or attributes are naturally associated with each other, such as the auditory perception of an increase and decrease in pitch is associated with visual or haptic perception of rising and falling in vertical position [15,45,36]. These kinds of associations can be found in other cross-sensory features as well, including the sound-brightness association, sound-quantity association etc.[52,15,40,18]. As such, the term crossmodal correspondence (CC) in this paper refers to 'a compatibility effect between attributes or dimensions of a stimulus (i.e., an object or event) in different sensory modalities (be they redundant or not)'[45]. Accumulating behavioural studies show that when values of these associated attributes are combined congruently, for example increased pitch combined with increased vertical elevation, the accuracy and efficiency of people's sensory-motor response will be improved (for a detailed review, see [45]). However, if the attributes are combined incongruently (i.e. the direction of change of one of the involved attributes is reversed (reversed polarity)), for instance, when increased pitch is associated with downward visual stimuli, people's performance will be lowered.

Some research studies have pointed out that many CCs may occur due to the repeated exposure to the natural environment, from which people learned perceptual regularities of associated crossmodal physical features [45,46]. Praise et al. empirically investigated the statistical mapping between sound frequency and the perceived elevation of the sound source [36]. With a wearable two directional headphone, a large sample of natural sound recordings was collected by participants who move freely indoors and outdoors. The analysis of recordings reveals a consistent mapping between the sound frequency and the physical elevation of the source of the sound. Furthermore, a consistent observation concerning the frequency-elevation correspondence was found when participants were asked to do a sound localisation task. Even though the correlation between the statistical environmental mapping and the perceptual judgement performance cannot be fully explained for now, the above investigation provides some experimental grounds for a linkage between CC perception and action [2,46].

Recent research also suggests that perception of crossmodal information can be influenced by task-irrelevant contextual features [54,8]. Walker and Walker conducted an experiment to address the relative perception of crossmodal feature values. In their experiment, participants were presented with six circles with different luminance. Three of the circles were brighter than the background colour, and the other three appeared darker. Participants were asked to classify whether each of the circles was brighter or darker than the background, and confirm their answer by pressing one of two differentlysized keys. Results showed that participants classify brighter circles more quickly with the smaller key, and classified darker circles more quickly with the larger key. This finding suggests that people's crossmodal perception is not absolute, but that it can be subject to ambient environmental influence subliminally, which in turn modulate sensory-motor activities. Moreover, this finding concerning the relative values of the modal stimulus being studied, points to the need of investigating sequential activities in response to a range of values of stimuli, rather than focusing only on effects induced by values at the extreme ends of a range, which were validated through the task paradigm involving one-shot actions or one-time judgements [15,45]. The findings of such studies might be argued to have limited applicability to real-world interactions, as real-world tasks tend to have ranges of multimodal signals and often involve a series of interactions over time.

### 2.2 A view from HCI design implementation

From an HCI perspective, we are aware of two studies that have investigated the design possibilities of CCs employing interactive settings requiring engagement over a longer period of time and involving a range of stimuli. Metatla et al. evaluated crossmodal interaction performance with different congruency levels between visual and audio modalities [31]. They conducted their experiment with a modified classification paradigm, which adopted a game mechanism and required a sequence of inputs when responding to the unimodal and multimodal stimuli. Results showed that the effect of crossmodal congruency between shape and elevation was confirmed in the visual display condition, but not in the multimodal conditions. Furthermore, the audio feedback cancelled the effect of crossmodal congruency levels, which implies that participants tended to rely on audio feedback to compensate when the interaction involved less congruent crossmodal feedback. In another experiment, Feng and Stockman tried to replicate the crossmodal effect through augmented physical features (of a tangible object) on an interactive tabletop [17]. Participants were required to discover a hidden target by moving the object around the table, which had concurrent augmented feedback indicating the distance between the object and the target. Visual, auditory and haptic modalities were combined into unimodal, bimodal and trimodal feedback. Results revealed that more accurate movement and efficient corrective actions were achieved with bimodal and trimodal feedback. However, the direction of five out of the thirty participants initial movements in response to crossmodal feedback during the trials was in the opposite direction to that which would be expected given their verbalized feedback in the post-experiment questionnaire.

This result suggested that people's sensory-motor reaction may not solely be determined by crossmodal stimuli in a bottom-up manner, but sometimes rather influenced by previous experience and current interaction goals.

From a human-centred design perspective, researchers have discovered and implemented self-activated crossmodal associations through an iterative design process. Bakker et al. applied a human-centred iterative design approach during several music education workshops, through which audio-haptic crossmodal associations have been identified from self-generated movements, such as increased volume being reflected by increased speed of movement and spatial elevation [4]. Through iterative design, the identified crossmodal associations which were subsequently implemented on hand-held tangible musical interfaces. A subsequent music learning session showed that participants who worked with the crossmodal congruent interfaces were more successful in music reproduction tasks. Though the crossmodal associations identified from this design process require future verification through more systematic experimental investigation, this study suggests a possibility of eliciting crossmodal correspondences through a top-down design process, which improves subsequent interaction experience.

## 2.3 Subliminal cueing and cognitive priming in HCI

Differing from the cognitive priming described in the HCI research field, priming, by definition in the cognitive psychology, is a form of non-declarative memory which does not involve conscious recollection of knowledge or previous experience, but instead reveals itself through behaviour [16]. The priming can occur either perceptually or conceptually, when repeated presentation of a stimulus leads to improved cognitive performance of either its perceptual feature or its meaning [11,25,49]. HCI researchers investigated the possibility of leveraging priming for the purpose of facilitating interaction [1,24], as well as enhance cognitive performance [26,12,20], through a commonly used technique which has been referred to as cognitive cueing/priming.

Two major categories of existing priming techniques are subliminal cueing and cognitive priming. The technique of using subliminal stimuli to trigger fast and automatic responses is referred to as subliminal cueing, which adopts strict experimental control over the factors of cueing time and content. The cueing time is less than 50 ms, within which the perception is assumed to be subliminal, and the cueing is commonly an external stimulus presented as a visual image or geometric shape, which exposes information that would appear during subsequent tasks. This technique was inherited from the masking paradigm from experimental psychology by HCI researchers, for the purpose of investigating the effect of priming on selection behaviour [3,38]. Thus subliminal cueing is also known as subliminal visual masking. Aranyi et al. conducted an experiment with subliminal visual cues preceding an object selection task in a virtual environment. A short-lived (within one second) impact of the visual cues on subsequence selection behaviour was observed [3]. Recently, a similar masking paradigm with three types of visual cues has been applied in a mobile application. Evidence showed that the subliminal stimuli with the particular time window of 17ms of this paradigm were not fully subliminal to the participants [38]. In summary, investigations of cueing effects using the masking paradigm, employing visually-dominated stimuli with less than a 50 ms duration, seem to have variable results in modulating interaction behaviour and the reliability of the results needs to be further tested. Therefore, this cueing technique is not suitable for tackling the present research questions.

The technique of cognitive priming has been applied with priming material either in the form of visual images, video, or textual stories [26,24,20]. The material does not always have a direct relationship with the interaction tasks and it can be presented either before or during interaction with no time duration limits. The purpose of this technique is to enhance cognitive function and affective state by presenting an implicit priming material in a top-down manner. Harrison et al. employed text-based stories as the priming material to investigate their influence on subsequent visual judgment performance on different types of charts. Results showed that positive priming improved participants'visual judgment accuracy [20]. The effect of cognitive priming could be observed even after the priming stage. Lewis et al. applied affective computational priming, in the form of background pictures, in a creativitysupported design tool [26]. With the concurrent priming material, this led to an improvement in the quality of participants creativity, but did not change the quantity of material they created. The priming effect remained even when the priming material was removed in the posterior task [26]. In conclusion, in the existing HCI literature, the reliability of applying the masking technique needs to be further tested. In comparison, cognitive priming has been used to invoke previous experience and mental states during goal-oriented tasks. Following this line of investigation, in the present study, we used a cognitive priming technique to invoke crossmodal experience and measured its effect on subsequent sensory-motor response.

**Table 1** Experimental design for experiment 1.

| | Manipulation groups | | | | Control groups | |
|---|---|---|---|---|---|---|
| Conditions | Condition 1 | Condition 2 | Condition 3 | Condition 4 | Condition 5 | Condition 6 |
| Priming factor | P-prime for pitchbrightness | C-prime for pitchbrightness | P-prime for pitchelevation | C-prime for pitchelevation | Control group for pitchbrightness | Control group for pitchelevation |
| factor | Pitchbrightness mapping | Pitchbrightness mapping | Pitchelevation mapping | Pitchelevation mapping | Pitchbrightness mapping | Pitchelevation mapping |

## 3 Research aim and scope

Previous studies have investigated crossmodal effects in response to changes in physical values that were varied either congruently or incongruently, where each of the variables was represented in one of two modalities. The changes to the parameter values of the stimuli have usually been between two polarized values of the parameters concerned [45,8]. However, crossmodal stimuli switching between extreme values in each of their respective modes, matched either congruently or incongruently, may rarely be encountered in everyday interactive environments. More often the question is first, whether and how people attend to available crossmodal information which has mediated feature values; and second, how people integrate several crossmodal information streams that are congruent when separated but which may be congruent or contradictory to

each other when combined? Specifically, the present paper seeks to address the following research questions:

RQ1: whether peoples crossmodal perception can be enhanced by contextual information, i.e. the priming technique, and if it can be, to what extent it modulates small and fast sensory-motor responses in an interactive, goal-oriented task? Experiment 1 was designed to tackle this question by using different types of priming material as a manipulation factor. In a goal-oriented task, we postulate that if people's perceptual preference is solely invoked by congruent crossmodal stimuli, people's performance will be equally good due to the CC effect, whether priming is present or absent [45,14]. However, if the process is not fully bottom-up, but can also be enhanced through top-down cognitive priming, there would be a perceptual reinforcement of correlated crossmodal information. Therefore, an improved performance should be observed during the subsequent task.

RQ2: How do people integrate crossmodal information in which two CCs are combined where the perception of the congruency of one CC excludes the perception of the congruency of the other? Furthermore, how, if at all, does the integration of such information change in the presence or absence of cognitive priming? Experiment 2 was designed to tackle this question by employing cognitive priming, we could deduce the integration process by observing whether the perception and corresponding task performance with primed crossmodal information is enhanced or reduced. If the perception of primed stimulus is distracted by another unprimed crossmodal stimulus, it is likely that the crossmodal cues have all been taken into account additively [10], thus the subsequent performance should be lowered by the distractor cue. If the perception of the primed stimulus is enhanced by the priming material, attention is likely drawn to the correlated crossmodal information selectively [10], i.e. people are less susceptible to the distractor cue, thus the performance in the subsequent task should be better.

There are two CCs and two types of priming materials involved. Based on previous empirical findings concerning CCs, we chose 2 pairs of crossmodal stimuli that have been frequently referenced in the literature for our experiments: the pitch-brightness and the pitchelevation crossmodal mappings [15,45,36]. We also introduced two types of priming as the manipulation factor. One type used bottom-up stimuli, i.e. physical features of brightness, elevation and pitch, to imply the CCs, namely the perceptual priming; and another type used top-down stimuli, i.e. videos in which the brightness, elevation and pitch were embeded with meaningful contents, namely the conceptual priming. Conceptual priming employs a crossmodal stimulus in the form typical of the natural environment [36], while perceptual priming emphasises the physical features or values of the stimuli. Specifically, the conceptual priming context uses naturalistic sound and corresponding video clips to represent the correspondence between auditory pitch and visual brightness, as well as auditory pitch and visual elevation [36,2]; while perceptual priming uses musical notes with a set of congruent visual icons to implement the same CCs. Further details about the priming materials will be explained in the apparatus section of the paper.

**4 Methods**

The same design and procedure were used for two experiments. In this section, we report the general experimental protocol, followed with experiment 1. Then in section 6, we explain the change in the manipulation factor and report experiment 2.

### 4.1 Experimental design

The experiments have to be a between-subjects design to eliminate the learning effect of priming. There were two independent variables. The first was the type of priming: the perceptual priming (P-prime) and the conceptual priming (C-prime). The second variable was the two crossmodal mappings, the pitch-brightness correspondence and the pitch-elevation correspondence. The two experiments have 6 conditions with 4 manipulation groups and 2 control groups. The manipulation groups were: P-prime and C-prime on pitch-brightness mapping respectively (condition 1, 2), the P-prime and Cprime on another crossmodal mapping, pitch-elevation respectively (condition 3, 4). The control groups were pitch-brightness mapping without priming (condition 5) and the pitch-elevation mapping without priming (condition 6). The manipulation factors and experimental conditions are listed in table 1.

### 4.2 Task and general procedure

The task in both experiments is learning and reproducing crossmodal sequences. The sample sequence was made up of two parts, the auditory melody with five pitch values equal to musical notes C, D, E, G, A; and the visual counterpart with either five levels of brightness or vertical position displayed in five circles on the screen (Figure 1 Experiment 1a). The detailed information about crossmodal mappings is presented in the experimental platform section. During each trial, a randomly generated sample sequence was displayed once, which has melody matched with the concurrent visual stimuli. Participants are required to reproduce the melody by clicking the circles on the screen to reproduce the melodic sequence.

The general procedure is as follows. First, participants were given a consent form with details of the experiment to read and sign. Then they were introduced to the priming session based on which group they were assigned. Participants who were in manipulation groups were told to watch a short video or an animation, while control groups had no priming session and simply moved to the next session directly (Figure 2). After priming, participants move to the warmup session to get familiar with the task procedure and the hardware setup. To ensure everyone received practice without overtraining, they were instructed to practice no less than twice and no more than six times. The crossmodal stimuli used in the warm-up session were different from those in the task session, for the purpose of minimising learning effects. Specifically, the sequences in the warm-up session used three-note sequences, while the task session used five-notes sequences. After practice, participants moved to the task session, which contained 16 trials (Figure 2). All the sessions were programmed in a single piece of software, participants moved to the following section by pressing a next button on screen.

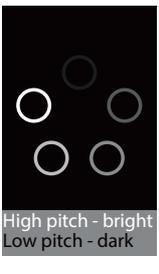

| Exp | Experiment 1 | | Experiment 2 |
|---|---|---|---|
| | a | b | Pitch - brightness vs. Pitch - elevation |
| CC | Pitch - brightness | Pitch - elevation | |
| Interface layout | 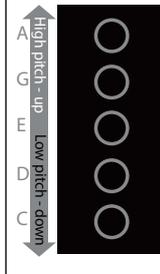 | | 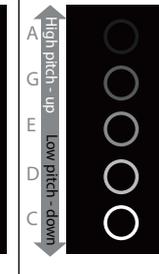 |

| | HSL display | 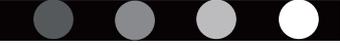 | | | | |
|---|---|---|---|---|---|---|
| Visual | Hue | 0 | | | | |
| | Saturation | 100 | | | | |
| | Lightness | 12 | 34 | 56 | 78 | 100 |
| Audio | Notes | C4 | D4 | E4 | G4 | A4 |
| | Frequency | 261.63 | 293.66 | 329.63 | 392.00 | 440.00 |

**Fig. 1** Interaction trial layout. Experiment 1a implement the pitch-brightness mapping, the bright circle paired with high pitch sound and the dark circle paired with low pitch sound. Experiment 1b implement the pitch-elevation mapping. The arrow indicates crossmodal mapping directions. Experiment 2 implement the two crossmodal mappings arranged in a mutually exclusive manner.

After participants finished all the trials, they were asked to complete a post-experiment questionnaire. The first section of this questionnaire collected participants demographic information, music training history, and whether they had any visual or auditory disorder recently. The second section collected subjective evaluation data, including their interpretation of the priming material, and the interaction strategy they used during the game. The entire experiment lasted for 7 to 13 minutes.

compared with 'dark, heavy, gloomy' things that tend to be observed near or on the ground [39]. More recently, the results of empirical investigations [45,36, 44] suggest that some CCs do indeed arise from the experience of correlated physical properties in the natural world. Hence, the conceptual priming in the present study was designed based on naturalistic sound and visual representations. Instead of using an image or picture as the priming material, which contains only visual information [3,37], we used video or animation to display crossmodal information. For the pitch-brightness crossmodal mapping, we composed sound clips of birds singing and land-based carnivores roaring to represent high pitch sound and low pitch sound respectively; and in the visual mode, a mute time-lapse video of night-today was used to represent the change of brightness (Figure 3 a). This approach was developed based upon the sensory experience that birds are usually active in the daylight and land-based carnivores are usually active at night, thus high pitch sounds tend to be associated with brightness and low pitch sounds with darkness. The composed video are available to review from (Hyperlink: Pitch-brightness) and from ( Hyperlink: Pitchelevation).

For the pitch-elevation crossmodal correspondence, we used the same sound clips to represent the high and low pitch, and composed a mute video of birds flying in the sky and carnivores running on the ground to represent the vertically high or low elevation (Figure 3 b). The sensory experience that we build upon is that high pitch sound corresponds to high in the sky and low pitch sound corresponds to being on the ground [39,45].

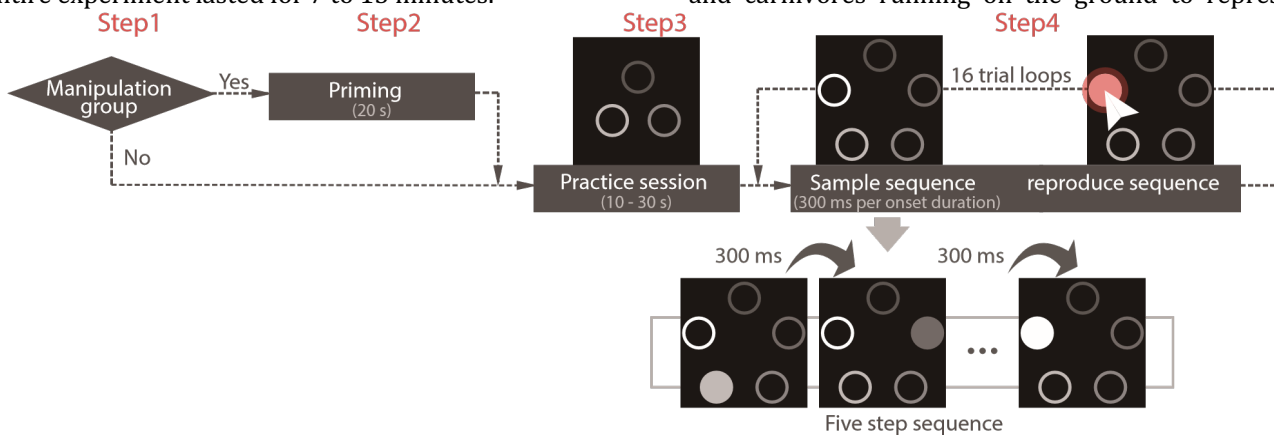

**Fig. 2** Experimental procedure. Step 1, assign groups; Step 2, manipulation groups will watch priming video or animation, and control groups will go to S3 directly; Step 3, practice; Step 4, do 16 trials with each one contains a 5 step sequence to learn.

### 4.3 Apparatus

#### 4.3.1 Conceptual priming material

Early studies claim that our experience of nature can form the basis of a CC. For example, 'thin, small, light and airy' things tend to be found at relatively high altitude,

#### 4.3.2 Perceptual priming material

Based on cognitive studies of crossmodal correspondence [45,36], we used piano sample sounds to represent pitch values. For the priming of pitch-brightness, the notes were presented in the order C-D-E-G-A-A-G-ED-C, and were paired with a circle which has synchronized changing in brightness from dark to bright and back to dark. The same sound samples with the same playing sequence were used in the priming for the pitchelevation matching, with a synchronized visual display of a circle moving from the bottom to the top on the same background.

### 4.3.3 Experimental platform

The platform for experiment 1 has the pitch-brightness and the pitch-elevation mapping arranged separately in different conditions, thus both of the stimuli are congruent without interfering with each other e.g. increasing pitch corresponded with upward position or brighter visual appearance (Figure 1, Experiment 1a and 1b). The audio was presented at a constant volume level of 6 (on Mac Pro device, which was tested for comfortable hearing) through a WH-CH500 headphone.

As a screen-based study [50,28], the HSL (hue-saturationlightness) colour scheme was used in the visual representation. We keep the H = 250, S = 0, and the lightness value ranging from 12 to 100 with the interval of 22. The corresponding auditory stimuli were notes C, D, E, G, A. Each of the five notes lasted for 300ms with 100ms intervals in-between. All the elements were displayed on a black background (H = 0, S = 100, L = 0) with a screen resolution of 2560*1600 dpi. In order to avoid the potential confound of stimulus-response compatibility effect [40] (e.g. high pitch naturally correlated to up/right and low pitch to down/left), instead of a linear arrangement, the circles were organized in a pentagon formation (Figure 1 a), during each trial a randomly generated sequence is displayed. The position of each circle placed in the pentagon was also randomised.

### 4.4 Measurements

Two depended variables were time intervals between each input and task error rate. Time intervals were counted as the length of time between adjacent clicks. The difference in time intervals between the manipulation groups and the baseline groups was calculated as an indicator of the priming effect on sensory-motor reaction efficiency. Interaction error rate was calculated by dividing the number of wrongly produced notes by the total number of sample notes.

In addition, the qualitative data that was included in the analysis was as follows: 1) the subjective interpretations of priming material collected from the postexperiment questionnaire, and 2) Plots of the overall task accuracy which was calculated not by summing the number of wrongly produced notes, but by the number of incorrectly alined CCs during reproduction. For example, if one step in a sample sequence change from note C to note E with the brightness going up 2 levels, both the reproduced move from C to E with brightness going up 2 levels and D to E with brightness going up 1 level would be counted as correct, since the increased pitch value corresponded with brightened visual display. However, the step from A to E with brightness level going down would be counted as an incorrect step, as the decreased pitch sound should not be aligned with the increased brightness level.

### 4.5 Hypotheses

Since the pitch-brightness and pitch-elevation correspondences have been repeatedly evaluated in the literature as working for most people [15,40,18], we assumed that most participants would be able to do the task without perceptual discrepancy on those crossmodal stimuli. Thus we hypothesized that:

H1: Following previous studies [34,24,26], we predict that people's crossmodal perception can be enhanced by cognitive priming. Specifically, that both the

| Con | Visual priming | Auditory priming | CC implied |
|---|---|---|---|
| a.Condition1and2 | 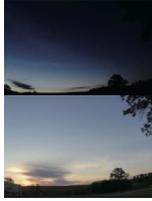 Video frames of night to day | 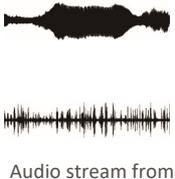 Audio stream from carnivore roaring to birds singing | Pitch-brightness correspondence<br><br>Night to day implies the change of visual brightness level, carnivore roaring to birds singing implies the change on pitch levels |
| b.Condition3and4 | 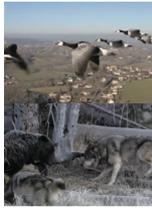 Video frames of sky and ground | 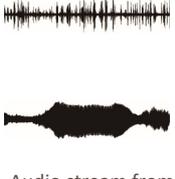 Audio stream from birds singing to carnivore roaring | Pitch-elevation correspondence<br><br>The view from sky to ground implies the change of visual elevation, birds singing to carnivore roaring implies the change on pitch levels |

**Fig. 3** Video screenshots and partial waveform intercepted from conceptual priming material for pitch-brightness crossmodal mapping (a) and pitch-elevation mapping (b).

P-prime and the C-prime for the pitch-brightness mapping and the pitch-elevation mapping will support faster sensory-motor responses and produce better task accuracy than would be seen in the two baseline groups.

H2: The type of priming will have different sensorymotor modulation effects on task performance. However, we chose not to predict the direction of the difference.

## 5 Experiment 1: Study on congruent crossmodal mappings

One hundred and twenty participants (67 male, 53 female, aged 18-55 years, mean = 26.17, SD = 5.56) were involved in the first experiment. All participants confirmed that they have no visual and auditory disorders before trials (after correction). Participants were balanced across groups according to their age and gender. In order not to introduce a potential confounding factor of cultural background on CC perception or interpretation, a mixture of volunteers of different nationalities, professions and music training experience were recruited through the universities' e-mail list and social network sites.

**Table 2** Statistical analysis based on time intervals in experiment 1.

one-way

| | F(2, 3837) | p | r | Note |
|---|---|---|---|---|
| Pitch-brightness mapping | 76.649 | .000 | .17 | P-prime (517.57 ms) < Control group (555.04 ms), C-prime (500.70 ms) < Control group |
| Pitch-elevation mapping | 22.246 | .000 | .10 | P-prime (578.43 ms) < Control group (558.48 ms), C-prime (545.98 ms) < Control group |

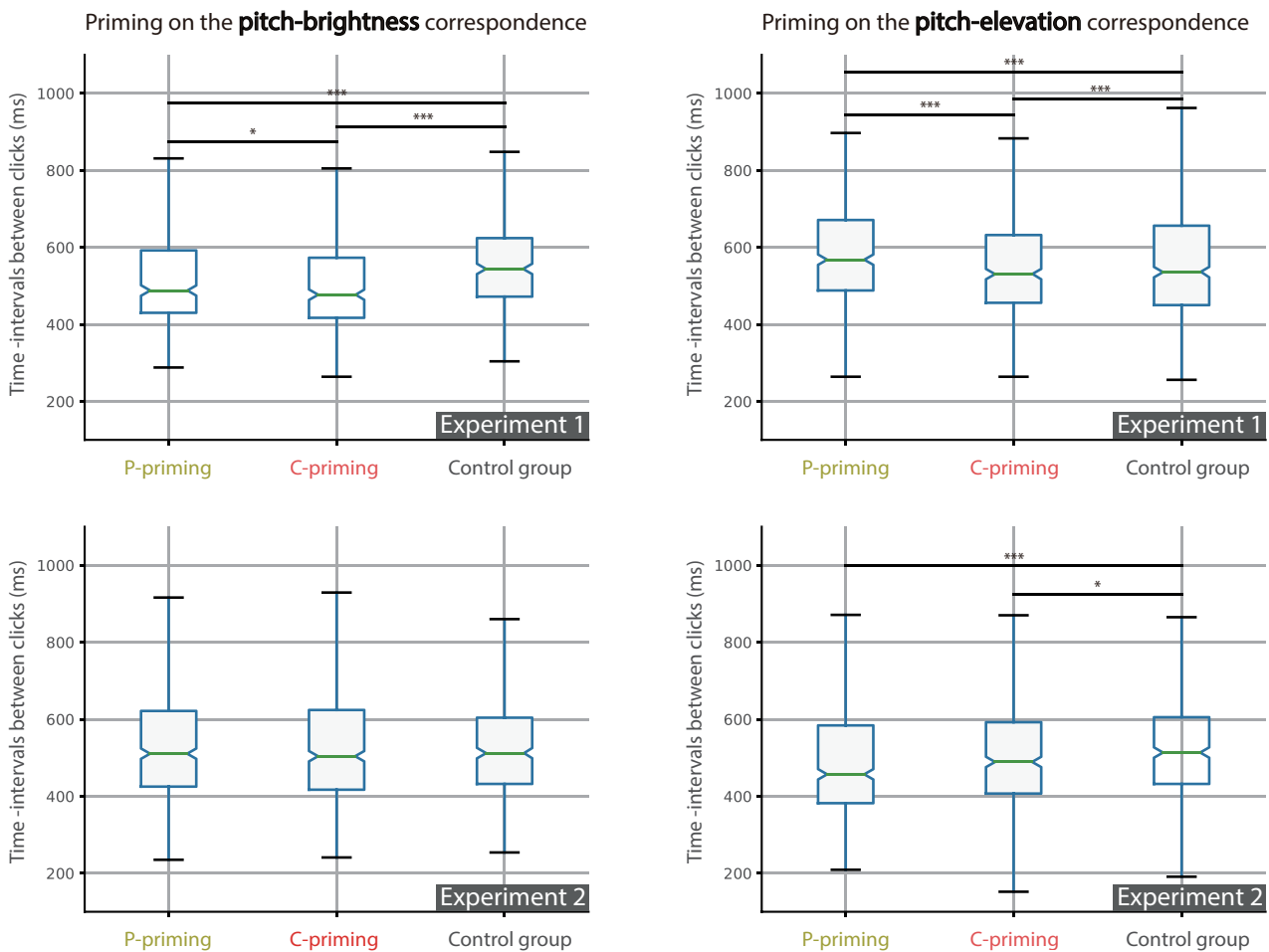

**Fig. 4** Box plot of the time intervals under the priming for the pitch-brightness correspondence and for the pitch-elevation correspondence in experimet 1(top two panels) and experiment 2 (bottom two panels). ∗$P < 0.05$, ∗∗∗$P < 0.001$.

## 5.1 Results

A Kolmogorov-Smirnov test showed that the time intervals data can be assumed to be normally distributed. We ran a ANOVA to compare the results of Pprime, C-prime and baseline conditions on pitch-brightness mapping, as well as the P-prime, C-prime and the baseline condition on pitch-elevation mapping. The Fisher's LSD test was used for post-hoc tests of main effects. We used a confidence level of $\alpha$ = 0.05 for the tests.

*5.1.1 Results on time intervals between inputs of a sequence*

During experimental trials, two of the priming groups with the pitch-brightness mapping had statistically significantly smaller time intervals between clicks than the control group, and the C-prime group also produced smaller time intervals than the P-prime group (Figure 4). However, for the pitch-elevation mapping, only the C-prime groups had statistically significant faster inputs than the control group, while the P-prime group produced statistically significant slower inputs than the

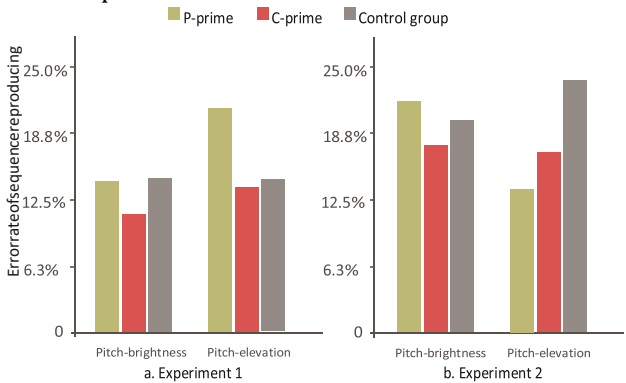

**Fig. 5** Error rate of sequence re-producing. Figure a shows results of error rate of experiment 1. Figure b shows results of error rate of experiment 2.

control group (Figure 4). The detailed statistical results are listed in table 2

*5.1.2 Results on task error rate*

For the pitch-brightness mapping, C-prime group (11.13%) produced more accurate sequences than the P-prime group (14.25%) and baseline group (14.44%) (Figure 5 a (pitch-brightness)). For the pitch-elevation mapping, the P-prime group (21.44%) produced a higher error rate than the control group (14.50%) and C-prime group (13.88%) as shown in figure 5 a (pitch-elevation).

### 5.2 Discussion

The hypothesis H1 that crossmodal perceptual preference can be induced by cognitive priming has been confirmed for the pitch-brightness correspondence. With the pitch-elevation correspondence however, only the C-prime group showed a positive effect on task performance, while the P-prime group had slower motor responses (Figure 4 (up-right chart)) and poorer accuracy (Figure 5 a (pitch-elevation)).

These inconsistent outcomes between the two crossmodal correspondences may result from the task paradigm an increase in response times and limited working memory capacity. This assumption can be supported by subjective evaluation on priming material, which will be discussed in the later part of the section. However, such degraded performance was less evident in the P-prime group with the pitch-brightness correspondence. The self-reports from the post-experiment questionnaire reflected that the interaction strategy used by participants may compensate the inferior effect. In response to a question about interaction strategies for reproducing the sequences, 67% of participants in the pitchbrightness condition said that they visually tracked the flash pattern of the sample sequence as a trajectory, which was easier to recall and repeat. In contrast, none of the participants in the pitch-elevation mapping group reported the same or a similar strategy. Indeed, the visual pattern for the pitch-elevation correspondence has a 1-D arrangement with an overlapped moving path, while the visual pattern for the pitch-brightness correspondence has a planar arrangement. Since visual perception is more sensitive to spatial arrangement, while auditory perception is more sensitive to temporal information, the combination of a 2-D visual stimulus with a 1-D auditory stimulus can produce a multisensory enhancement effect [23]. As a result, the time delay and the error were less salient in the conditions with a pitchbrightness mapping.

The hypothesis H2 that priming types will have different effects on sensory-motor modulation has been confirmed. Results showed that the C-prime condition did improve participant's crossmodal perception, in both the pitch-brightness and the pitch-elevation correspondences, which in turn supported faster motor response as well as improved accuracy of crossmodal sequence reproduction. In comparison, the P-prime condition had a positive effect only on the pitch-brightness crossmodal mapping, and the positive effect may be due to the compensatory interaction strategy afforded by the interface. Previous studies and participants interaction strategies. Previous behavioural making [34,24], thus we deduce that the perceptual en-have shown that the priming process without conscious awareness can have a positive effect on peoples affective reaction and decision research studies mainly used a speeded classification paradigm [45], which required participants to react to stimuli with a one-shot key press. This process requires little working memory or cognitive resource. In the current experiment, however, participants had a clear interaction goal and engaged with a sequence of input actions. The P-prime, which primed for a direct association between the auditory and visual stimuli, possibly functioned as an explicit instruction, and encouraged participants to constantly recall and check during the interaction. This extra cognitive load may have led to hancement with C-priming in our case could also have functioned in a subliminal way. It enabled participants to have a stronger perceptual alignment with a specific CC, which facilitated fast and instantaneous motor response. While the P-prime material may rise to the level of conscious awareness during the priming process, thus involving working memory allocation while doing the task, which lowered the overall performance to some extent.

A supportive finding of the above assumption can be gleaned from the post-experimental questionnaire. Most of the participants in the C-priming groups were not aware of the CC presented during priming, and could not consciously recall the correlation between the priming and the interaction task. In all of the 40 subjective interpretations from the C-priming groups (2 Cprime groups with 20 participants in each), only 2 participants reported that 'it was about how the sound changes with visual information', and that 'they are there to focus your concentration on the screen and the sound'. The other 28 participants' answers were exclusively focused on the detailed contents of the prime context. Such as 'Its showing the sunrise' for the pitchbrightness priming and 'flying birds, running animal' for the pitch-elevation priming. Meanwhile, only 4 of the 40 participants rated that the C-prime was helpful for the interaction task. In contrast, 36 of 40 participants in the P-priming groups were fully aware of the purpose of the priming. One of the typical answers, for instance, was that 'information that helps you prepare for the game' for the pitch-brightness priming, and that 'different notes of music vertically spaced' for the pitchelevation priming. 36 of the 40 participants rated that the priming was helpful. From the usability perspective, this evaluation can explain the assumption of extra cognitive load that was discussed for the hypothesis H1.

Combining participants' behavioural data with their subjective rating, we confirmed that the C-prime induced enhanced perception and led to better task performance, even when the whole process was subliminal. This fact also explains why the C-priming was not considered helpful by most of the participants. In comparison, the purpose of the P-priming was recognised in most cases and thus tended to be acknowledged to be helpful for task completion, however, the actual behavioural data of the P-primed participants pointed to the opposite.

To better understand the quantitative effects of the priming techniques on subsequent goal-oriented tasks, participants performance accuracy based on crossmodal alinement, as explained in section 4.4, was plotted on two temporal scales: the within-trial performance and between-trial performance (Figure 6 Experiment 1). Along the horizontal axis, which represents the steps in reproducing the sequence during trials, the P-prime groups produced more accurate moves in the first two steps and were more prone to error in the last two steps. In comparison, the C-prime group and the control group do not have a similar pattern along the horizontal axis. These observations reflect that participants in the Pprime group had different interaction strategies, which relied more on working memory capacity than on crossmodal congruency. This observation is consistent with participants behavioural data discussed above. Along the vertical axis, which represents the between-trial performance, no improvement can be observed in the later trials due to practice in all groups, as a matter of fact, the C-prime groups produced more accurate performance in the first half of the total trials. The P-prime groups and the control groups also had slightly better performance in the first few trials, but the difference was not as obvious as that in the C-prime groups. These observations reflect that the effect of cognitive priming on crossmodal perception had a more noticeable effect in the early stages of interaction, especially in the case of C-priming, but it did not persist into the later stages.

# 6 Experiment 2: Study on mutually exclusive crossmodal mappings

One hundred and twenty participants (62 male, 58 female, age 18-35 years, mean = 26.48, SD = 4.58) were recruited for the second experiment, with 20 participants in each condition. All the participants were newly recruited for experiment 2, and randomised in the same way as in experiment 1.

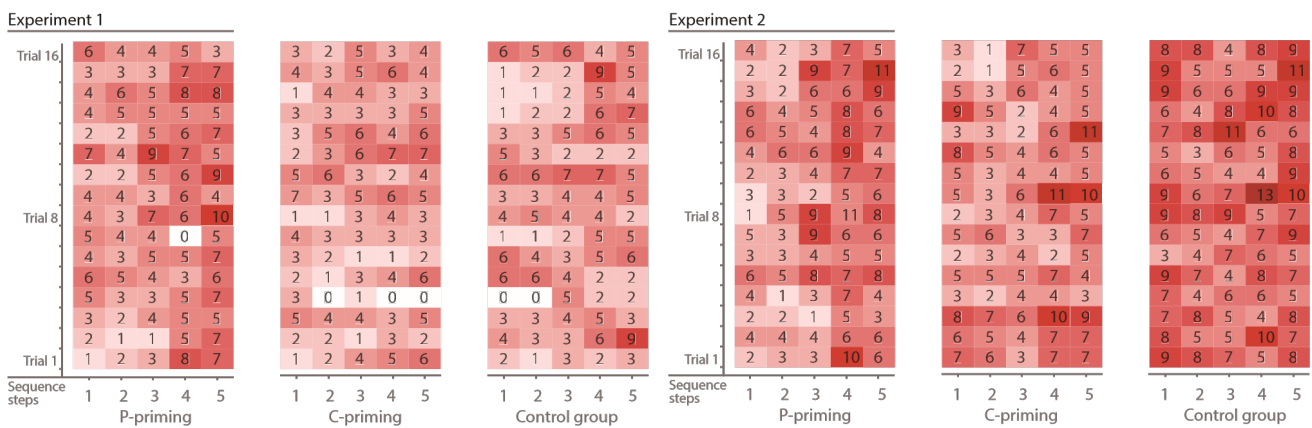

**Fig. 6** Crossmodal alignment in the P-prime group, C-prime group and the control group of experiment 1 (left) and 2 (right). The horizontal axis denotes the sequence steps from 1 to 5 (within trial), and the vertical axis denotes the number of experimental trials (between trials). The digits with the red background represent the number of inputs which were not aligned with CCs at that specific step (counted with all participants' steps in each group).

**Table 3** Experimental design for experiment 2.

| | **Manipulation groups** | | | | **Control groups** | |
|---|---|---|---|---|---|---|
| Conditions | Condition 1 | Condition 2 | Condition 3 | Condition 4 | Condition 5 | Condition 6 |
| Priming factor | P-prime for pitchbrightness | C-prime for pitchbrightness | P-prime for pitchelevation | C-prime for pitchelevation | Control group for pitchbrightness | Control group for pitchelevation |
| CC factor | Pitchelevation as a distractor | Pitchelevation as a distractor | Pitchbrightness as a distractor | Pitchbrightness as a distractor | Pitchbrightness mapping with pitchelevation as a distractor | Pitchelevation mapping with pitchbrightness as a distractor |

This experiment was designed to investigate the second question: RQ2: How do people integrate crossmodal information in which two CCs are combined where the perception of the congruency of one CC excludes the perception of the congruency of the other? Furthermore, how, if at all, does the integration of such information change in the presence or absence of cognitive priming?

Participants in all conditions were introduced to the interface which contained 2 crossmodal mappings arranged in an incongruent manner (Figure 1 c). Participants in conditions 1 and 2 were primed in relation to the pitch-brightness correspondence, which was applied in the subsequent task but has pitch-elevation correspondence as a distractor. Participants in conditions 3 and 4 were primed in relation to the pitch-elevation correspondence, with the pitch-brightness correspondence as the distractor (see the experimental design in table 3). In this way, participants in each of the manipulation groups can only be perceptually consistent with one of the CCs, but can not be consistent with both.

H3: For the 2 CCs arranged in a contradicted manner, the primed groups will integrate crossmodal information in a selective way, while the control groups will do in an additive way [10]. Specifically, the priming groups will produce better performance e.g. faster motor response and lower error rates than the control groups.

6.1 Results

A Kolmogorov-Smirnov test showed that the data set can be assumed to be normally distributed. The oneway ANOVA was used as statistical method with a confidence level of $\alpha = 0.05$.

### 6.1.1 Results on time intervals between inputs of a sequence

When primed on the pitch-brightness correspondence with pitch-elevation as an interaction distractor, there was no significant difference between primed groups and the control group, and there was no significant difference between the two types of priming (Table 2, and figure 4 (bottom-left panel)). When primed on the pitchelevation correspondence with pitch-brightness as an interaction distractor, both the primed groups performed significantly better than the control group. There was no significant difference between the two types of priming (Table 4, and figure 4 (bottom-right panel)).

### 6.1.2 Results on task error rate

With the CCs arranged incongruently, when primed on the pitch-brightness correspondence but distracted with pitch-elevation during the interaction, the C-prime group (17.88%) produced a slightly better performance in the error rate than the P-prime group (22.06%), and the control group (20.13%) (Figure 5 b pitch-brightness). When primed on the pitch-elevation correspondence with the pitch-brightness as the distractor, the P-prime group (13.69%) showed a lower error rate than the C-prime group (17.19%) and the control group (23.94%) (Figure 5 b pitch-elevation).

## 6.2 Discussion

Hypothesis H3 has been confirmed for priming on the pitch-elevation mapping but rejected for the priming on pitch-brightness mapping. The two priming groups and the control group for the pitch-brightness correspondence have no significant difference in terms of motor response speed (Figure 4 (lower-left)), and there was no than the control group in terms of both the motor response speed and sequence re-producing accuracy.

One explanation for this fact may be that the pitchelevation correspondence may have stronger perceptual weight [10] than the pitch-brightness correspondence. The perceptual experience of the pitch-elevation correspondence may be encountered more frequently in daily interactions than the pitch-brightness correspondence, thus the neural response to pitch-elevation stimuli may be stronger than that for the pitch-brightness correspondence [5,47]. This crossmodal perceptual weighting may not be easy to observe in situations where crossmodal stimuli are isolated from one another, such as in the case of experiment 1 and in many classificationbased tasks employed in previous studies [15,45,9]. While in the situation of experiment 2, where crossmodal stimuli were overlapped and incongruent, priming on the relatively stronger CC, pitch-elevation correspondence in this case, enabled participants weighting the dominant stimulus selectively in the subsequent activity. However, priming on the less dominant CC, the pitchbrightness correspondence in this case, seems to have had little or no effect on modulating subsequent activity. As a result, only those groups primed on the pitch-elevation correspondence produced improved performance, both on motor response speed and task accuracy. Thus we postulate that crossmodal stimuli integration may occur in a selective manner rather than an additive manner in the presence of priming[22].

Following the discussion in experiment 1, to observe how the priming technique influenced performance in an incongruent crossmodal situation, the crossmodal alignment accuracy of participants has been plotted in figure 6 (right). The plot shows that the P-priming groups produced more accurate moves at the first few steps in their sequences. Compared with the P-priming groups in

**Table 4** Statistical analysis based on time intervals in experiment 2.

|  | F(2, 3837) | p | r | Note |
|---|---|---|---|---|
| Prime on pitch-brightness with pitch-elevation as a distractor | .886 | .412 | .000 | P-prime (537.17 ms) < Control group (492.21 ms), C-prime (527.74 ms) < Control group |
| Prime on pitch-elevation with pitch-brightness as a distractor | 23.145 | .000 | .11 | P-prime (494.23 ms) < Control group (525.79 ms), C-prime (506.54 ms) < Control group |

obvious difference between the primed groups and the control group in terms of the accuracy (Figure 5 b). The other two priming groups for the pitch-elevation correspondence had significantly better performance experiment 1 (figure 6 left), a similar pattern appeared regardless of how crossmodal stimuli were combined. Based on this observation from both of the experiments, we can deduce that participants presented with explicit

P-priming material tend to do the task in a way that involved more working memory capacity. Different from the situation in experiment 1, the extra attention allocation made participants less susceptible to perceptual distractions in experiment 2. The plot also shows that participants in the C-priming groups produced better performance in the later half of total trials than the earlier half. Comparing with the Cpriming groups in experiment 1, which shows better accuracy in the earlier trials, it is plausible to attribute the improvement produced in experiment 2 to the practice effect more than to the priming effect. However, the plot for the control groups for experiment 2 does not have a salient observable difference between earlier trials and later trials due to practice. This observation could indicate two things. First, the C-priming material does have an effect on performance. Due to subliminal priming, participants may not attend to the intended CC at the first few tries, but gradually it becomes more influential on the primed CC as the trials go on. Second, without priming, participants are susceptible to being influenced by the crossmodal distractor that can happen to correspond with one of the modalities in another crossmodal stimulus. Thus, in the absence of priming, the integration of mutually exclusive crossmodal stimuli more likely happened in an additive manner than a selective manner [10].

# 7 General Discussion

The present study examined participants'crossmodal perception enhancement on performance, as well as the crossmodal integration process in the conditions with and without cognitive priming. The first contribution is introducing and evaluating the priming technique as a way to induce crossmodal experience and enhance perception of particular crossmodal information. When interacting with multisensory interfaces in a real world situation, any external stimulus could act as a primer to the system users, either it is an explicit symbolised stimulus or a subliminal stimulus imposed from the ambient system [48]. The current study is the first step to investigate the possible effect of external priming on both crossmodal perception and on the correlated fast motor response. The second contribution, to the best of our knowledge, is the first investigation into the way people integrate crossmodal stimuli with mutually exclusive congruency in the presence of priming information in the context of an interactive, goal-oriented task.

In general, the results of experiment 1 revealed that when two visual-audio CCs were isolated and did not interfere with one another, conceptual priming was successful in enhancing crossmodal perception of the primed CC subconsciously, which led to faster motor response and improved task accuracy. While perceptual priming operated in a more explicit manner. However, with goaloriented tasks, the explicit approach to priming could function as an instruction, which caused extra cognitive resource allocation to recalling and comparing between the priming and the task stimuli. The results reflected that this process actually diminished the task performance, though it was regarded as being helpful for interaction by most of the participants. The results of experiment 2 revealed that in the situation where two CCs have mutually exclusive congruency, the priming technique has little or no effect on the less dominant CC (pitch-brightness in this case). On the more dominant CC (pitch-elevation in this case), however, explicit priming enhanced perception and led to better task performance, e.g. faster motor response and better accuracy.

Extending previous studies with graded crossmodal stimuli and tasks involving a sequence of fast inputs [54,31], our finding further suggested that peoples interaction behaviour can not necessarily be predicted based on single polarised physical features. Individuals seem to possess varied perceptual weightings regarding CCs. This perceptual preference can be dynamically enhanced or inhibited by different types of priming according to the way crossmodal information is organised. We also propose, based on these findings, that with CCs that have mutually exclusive congruency, priming on the dominant CC will strengthen the perceptual response on that crossmodal stimulus, while priming of the less dominant CC will have little or no facilatory effect on fast and consecutive motor responses. Future studies are needed however, to verify and further explore the impact of multiple CCs in interactive environments, with either continuous feedback [17] or graded feedback stimuli. The outcomes of this line of research could be applied to the scenario where information needs to be overlaid in augmented reality or virtual reality, or where sophisticated interactive actions are required in situations imposing a high cognitive load.

In terms of potential applications, the present findings could contribute to the field of mindless technology [1,37] and multisensory interaction [33]. Mindless computing emphasises the interaction with the subconscious mental process, which is characterized as fast, automatic, and requiring little or no effort. Mindless technology makes use of this mental process, aiming to guide people towards intended interaction behaviour without annoying or distracting them. Previous research showed that both environmental features and the interactive system itself can impose a subtle influence on people's choice and judgement subliminally during interaction [1,3,37]. Building on theoretical accounts of CCs, the present study contributes to the field in that it shows cognitive priming techniques can also influence interaction involving consecutive motor responses. Implicit conceptual priming can be used to improve interaction efficiency when information shares only one CC feature, while when there are two or more streams of crossmodal information which have mutually exclusive congruency, explicit perceptual priming seems to have a stronger influence on facilitating interaction.

The present study also provides insights for the design of multisensory interactive systems [33], which can require a broad range of multi-input modalities and spontaneous access to ubiquitous information. One of the most discussed advantages of multisensory interaction is expanding peoples'cognitive capacity by presenting several streams of information through different sensory channels [35,56,23]. The design implication of the results of the present study are that the changing values of two or more streams of information which share a CC presented through the same pair of crossmodal sensory channels may inhibit information processing, such as the case in experiment 2, in which incongruent mappings between auditory pitch to visual brightness and to visual elevation diminished task performance. Possible solutions can be implemented either by isolating the information streams spatially or temporally to avoid confusion, or by choosing different pairs of crossmodal sensory channels to reduce distraction, such as using both the visual-audio channels and the visualhaptic channels.

One of the limitations of the present study is the small effect of the statistical analysis. The small sample size involved in each experimental condition may be one of the reasons. Further verification with larger sample sizes should be conducted in the future. Another limitation of the present investigation is that it does not indicate whether the cognitive priming effect and the perceptual weighting phenomenon would generalise to other CCs, specifically with varied crossmodal values and intensity of crossmodal components. Although it seems unlikely that the priming contexts in the present studies are the only two cases that have effects. The next step in this research would involve systematically testing the priming approaches in terms of materials, themes (i.e. interaction context), priming duration and persistence. Sensory channels for priming should not be limited to the audio and visual modalities. future work should expand the scope of the priming modalities and modality combinations to haptics, olfaction, and gustation etc.

## Acknowledgements